\documentclass[aip,reprint]{revtex4-1}
\usepackage{graphicx}
\usepackage{dcolumn}
\usepackage{bm}
\usepackage{color}

\renewcommand{\vec}[1]{\mathbf{#1}}

\newif\ifgraph

\graphtrue             %

\begin{document}
\title{
Cargo Towing by Artificial Swimmers}

\author{Debajyoti Debnath}
\affiliation{Department of Chemistry,Presidency University, Kolkata 700073, India}

\author{Pulak K. Ghosh} 
\affiliation{Department of Chemistry,Presidency University, Kolkata 700073, India}

\author{Yunyun Li}
\email{yunyunli@tongji.edu.cn}
\affiliation{Center for Phononics and Thermal Energy Science, School of Physics Science and Engineering, Tongji University, Shanghai 200092, People's Republic of China}

\author{Fabio Marchesoni}
\affiliation{Center for Phononics and Thermal Energy Science, School of Physics Science and Engineering, Tongji University, Shanghai 200092, People's Republic of China}
\affiliation{Dipartimento di Fisica, Universit\`{a} di Camerino, I-62032 Camerino, Italy}

\author{Baowen Li}
 \affiliation{Department of Mechanical Engineering, University of Colorado, Boulder, Colorado
80309, USA}

\date{\today}

\begin{abstract}
An active swimmer can tow a passive cargo by binding it to form a self-propelling dimer. The orientation of the cargo relative to the axis of the active dimer's head is determined by the hydrodynamic interactions associated with the propulsion mechanism of the latter. We show how the tower-cargo angular configuration greatly influences the dimer's diffusivity and, therefore, the efficiency of the active swimmer as a micro-towing motor.
\end{abstract}
\maketitle

Artificial microswimmers are active particles capable of autonomous propulsion  \cite{Schweitzer,Granick,Muller}. This phenomenon, which can be regarded as a biomimetic counterpart of cellular mobility \cite{bacteria}, is fueled by stationary non-equilibrium processes activated by the swimmers themselves. As a result of some built-in functional asymmetry \cite{Gompper,Sen_rev}, self-propelled particles harvest kinetic energy from their environment, by generating local (electric \cite{Sen_rev}, thermal \cite{Sano}, or chemical \cite{SenPRL}) gradients in the suspension medium (self-phoresis). A common class of artificial swimmers is represented by the so-called Janus particles (JP), mostly spherical colloidal particles with two differently coated hemispheres, or ``faces''. The different functionalization of their faces lends these swimmers a characteristic dipolar symmetry which makes their axial propulsion possible \cite{GrahamPRL,Adjari,Golestanian}.

\begin{figure}[tp]
\centering
\includegraphics[width=7.5cm]{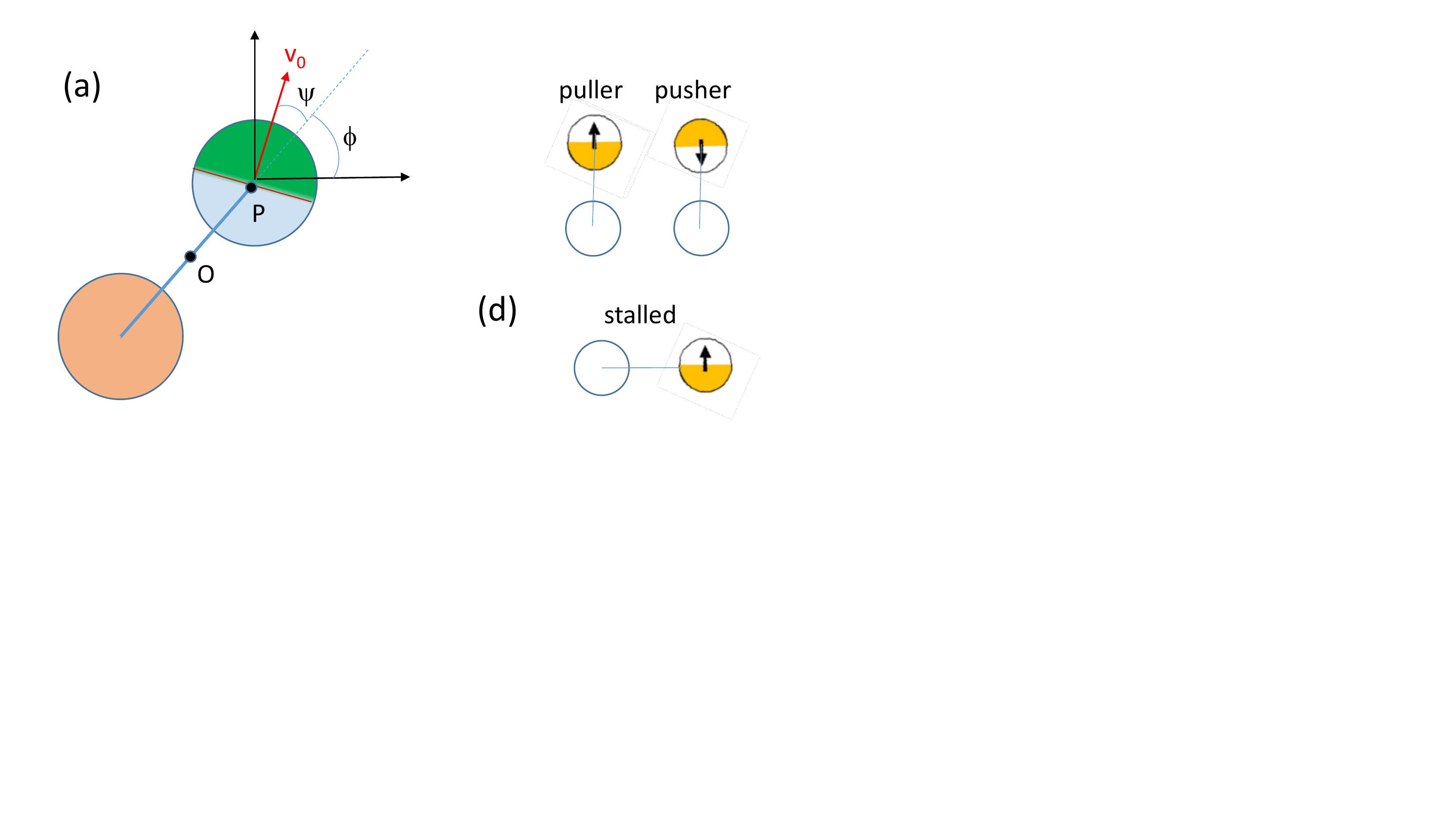}
\includegraphics[width=7.5cm]{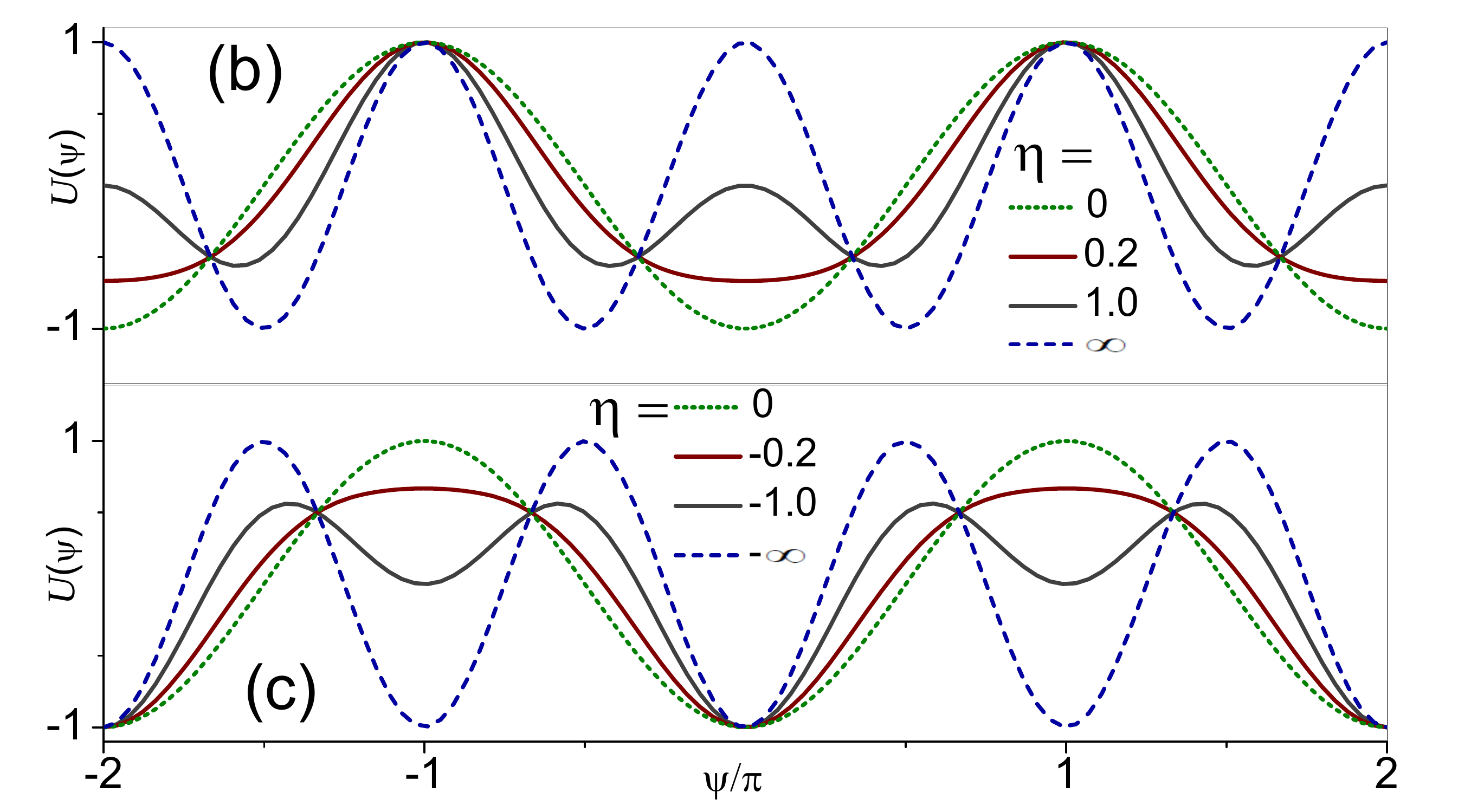}
\caption{(Color online) Active dimer's self-propulsion mechanism:
(a) the active dimer. $O$ and $P$ are respectively the center of mass and the center of force of a spherical Janus particle. ${\vec v}_0$ represents the instantaneous self-propulsion velocity vector; $\phi$ and $\psi$ denote the angle between the $OP$ axis and, respectively, the $x$ axis and ${\vec
v}_0$; (b)-(c) the angular potential $U(\psi)$, Eq. (\ref{U}), for different values of $\eta$, respectively, with $\eta>0$ and $\eta<0$; (d) dimer's limiting configurations for $\psi_0=0$ (puller), $\pi$ (pusher) and $-\pi/2$ (stalled), see text. \label{F1}}
\end{figure}

Among the prominent applications of artificial microswimmers and JP's, in particular, is their usage as motors, whereby they couple to a cargo, represented by a passive particle (PP), and tow it from a docking to an end station, often following a meandering path across a crowded environment \cite{small}. In such configuration, tower and cargo form a dimer with one active head, the JP, and a swerving tail, the PP \cite{Sano,GrahamPRL}. Biologists observed \cite{Lauga} that active swimmers can act either as pullers, with cargos trailing them, or as pushers, with cargoes positioned in front of them. Due to the hydrodynamic interactions between active and passive components, the JP and dimer axes are not necessarily aligned.
Such a short-range dipole interaction is associated with the self-phoretic  propulsion mechanism of the JP, independently of its radius, and decays quadratically with the distance from it \cite{Lauga}. As a consequence, we can assume that the dimer self-propulsion velocity fluctuates around one or more locally stable angles, $\psi_0$, with respect to the dimer axis. In this notation, $\psi_0=0$ ($\psi_0=\pi$) corresponds to the puller (pusher) configuration. Due to the cylindrical symmetry of such a bound system, more stable orientations of the propulsion velocity in the bulk come only by the pair, $\pm \psi_0$.

The purpose of this Communication is to propose a working model of active dimer's diffusion and, therefore, quantify the efficiency of JP's as cargo towers. We prove, both analytically and by means of numerical simulations, that varying the propulsion angle largely affects the dimer diffusion constant, with a maximum for the puller ($\psi_0=0$) and pusher ($\psi_0=\pi$) configurations, and a minimum, orders of magnitude smaller, in the stalled configurations with $\psi_0=\pm \pi/2$.

The average propulsion orientation of the dimer with respect to its axis is the result of two main competing mechanisms of hydrodynamic nature: (1) As the dimer's head tows its cargo, independently of its propulsion mechanism, at low Reynolds numbers a laminar flow is established,
which tends to align the dimer's axis with its propulsion velocity vector
${\vec v}_0$, the PP trailing the JP, i.e., $\psi_0=0$ \cite{Wurger,Geiseler,soft}; (2) Self-phoretic propulsion mechanisms acting upon the active head, are associated with short-range hydrodynamic backflows in the suspension fluid. Far from obstacles, the flow field in the vicinity of the JP is, to a large extent, axisymmetric around the JP axis. Such a symmetry is likely broken by a near bound cargo: the reaction of the constrained fluid backflow amounts to an effective torque, $f(\psi)$,  that tends to rotate the cargo around the active head \cite{force,walls1,walls2,walls3}; magnitude and sign of the torque depend on the surface properties of the dimer constituents \cite{Adjari,Wurger}.

We consider here a 2D active dimer, where both the center of force, $P$, and the center of mass, $O$, rest on its symmetry axis, see Fig. \ref{F1}(a). For simplicity, we assume that its constituents are (almost) pointlike spheres of identical mass and negligible radius, $a$, kept a fixed distance apart, $l$, with $a \ll l$. The dimer's moment of inertia can be set to one by introducing the dimensionless coordinates $x \to x/l$ and $y\to y/l$. Note that in dimensionless units the distance $OP$ is equal to 2 and at low Reynolds numbers the dimer's dynamics is insensitive to the value of bound masses. At a given time, $t$, the instantaneous self-propulsion ``force'' \cite{force} acting on the JP is oriented at an angle $\psi(t)$ with respect to $OP$ and fluctuates around the minima of an effective potential, $U(\psi)$, with constant modulus. In the overdamped regime, the resulting dimer's velocity, ${\vec v}_0(t)$, fluctuates parallel to the self-propulsion force, also with constant modulus, $v_0$. Moreover, the force applied in $P$ exerts a torque, $-v_0\sin \psi$,  around the center of mass, $O$, of the system. The resulting planar dimer's dynamics is thus modeled by a set of four Langevin equations (LE),
\begin{eqnarray} \label{modelLE}
\dot x &=& v_0 \cos (\phi+\psi) +\sqrt{D_0} \;\xi_x(t), \nonumber
\\\dot y &=& v_0
\sin (\phi + \psi) +\sqrt{D_0}\; \xi_y(t), \nonumber \\
\dot \phi &=& -v_0 \sin \psi +\sqrt{D_\phi}\;\xi_\phi (t), \nonumber \\
\dot \psi &=& f(\psi) +\sqrt{D_\psi}\; \xi_\psi (t),
\end{eqnarray}
where $\vec{r}=(x,y)$ are the coordinates of the dimer's center of mass, $O$, and $\phi$ the angle between its axis, $OP$, and the $x$ axis. The Gaussian noises $\xi_i(t)$, with $i=x,y,\phi,$ and $\psi$, are zero-mean valued, delta-correlated in time and statistically independent, that is, $\langle \xi_i(t)\xi_j(0)\rangle=2\delta_{ij}\delta(t)$. In the model Eqs. (\ref{modelLE}) we have implicitly assumed that the constants of the viscous forces acting upon the dimer's constituents were identical. Accordingly, upon breaking the bond connecting them, the JP would propel itself with speed $2v_0$ and both monomers, JP and the PP, would diffuse subject to independent translational noises, $\xi_x(t)$ and $\xi_y(t)$, of strength $D_0$. Finally, we remark that the condition $a\ll l$ allows neglecting corrections to the diffusive dimer's dynamics due to long-range hydrodynamic interactions \cite{Ermak}.

The torque $f(\psi)$ and the angular noise $\xi_\psi(t)$ appearing in the fourth LE (\ref{modelLE}), instead, model the effects of the short-range hydrodynamic interaction exerted by the tower on its cargo. For convenience, we express $f(\psi)$ in terms of a symmetric periodic potential, $U(\psi)$, that is,  $f(\psi)=-dU(\psi)/d\psi$.
A convenient choice for $U(\psi)$ is the angular function,
\begin{equation}
\label{U}
U(\psi)=\kappa_\psi[\eta \cos (2\psi) -\cos \psi]/(1+|\eta|),
\end{equation}
where $U(-\psi)=U(\psi)$, $U(\psi+2\pi)=U(\psi)$ and the $\psi$ domain can be restricted to $(-\pi, \pi]$. It has been demonstrated that a JP tends to move either inward or outward with respect to an obstacle, or even parallel to it \cite{walls1,walls2}, depending on the physical-chemical properties of the swimmer and the obstacle. We expect to observe a similar behavior also when the JP is bound to its cargo. For the sake of generality, we chose a tunable angular potential, $U(\psi)$, to reproduce any binding dynamics of the tower-cargo system. 

Indeed, as illustrated in Figs. \ref{F1}(b) and (c), the profile of the double cosine function of Eq. (\ref{U}) is governed by the parameter $\eta$ \cite{Sodano}. Special cases are the sinusoidal profiles corresponding to the configurations of Fig. \ref{F1}(d): (i) $U(\psi)=-\kappa_\psi \cos \psi$, with one stable minimum at $\psi_0=0$ (pusher), for $\eta=0$; (ii)  $U(\psi)=-\kappa_\psi \cos (2\psi)$, with two opposite stable minima at $\psi_0=0$ (pusher) and $\psi_0=\pi$ (puller), for $\eta \to - \infty$; and (iii)  $U(\psi)=\kappa_\psi \cos (2\psi)$, with two symmetric stable minima at $\psi_0=\pm \pi/2$, for $\eta \to \infty$. The transitions between the monostable configuration (i) and the bistable configurations (ii) and (iii) are marked by the appearance, respectively, of an additional local minimum, $\psi_0=\pm \pi$, for $\eta <-1/4$, and one couple of symmetric absolute minima, $\pm \psi_0=|\arccos (1/4\eta)|$, for $\eta >1/4$. Propulsion angles approaching $\psi_0=\pm \pi/2$ hint at a {\it chiral} dynamics \cite{Lowen,AoEPL}, where the JP tends to circle around the PP, thus reducing the ability of the tower to move its cargo (stalled configuration). Moreover, the amplitude of $U(\psi)$ varies between $2\kappa_\psi$ for $\eta=0$ and $(8/3)\kappa_\psi$ for $\eta \to \pm \infty$, meaning that the strength of the angular coupling between the two monomers is actually controlled by the coupling constant $\kappa_\psi$.

\begin{figure}[tp]
\centering
\includegraphics[width=7.5cm]{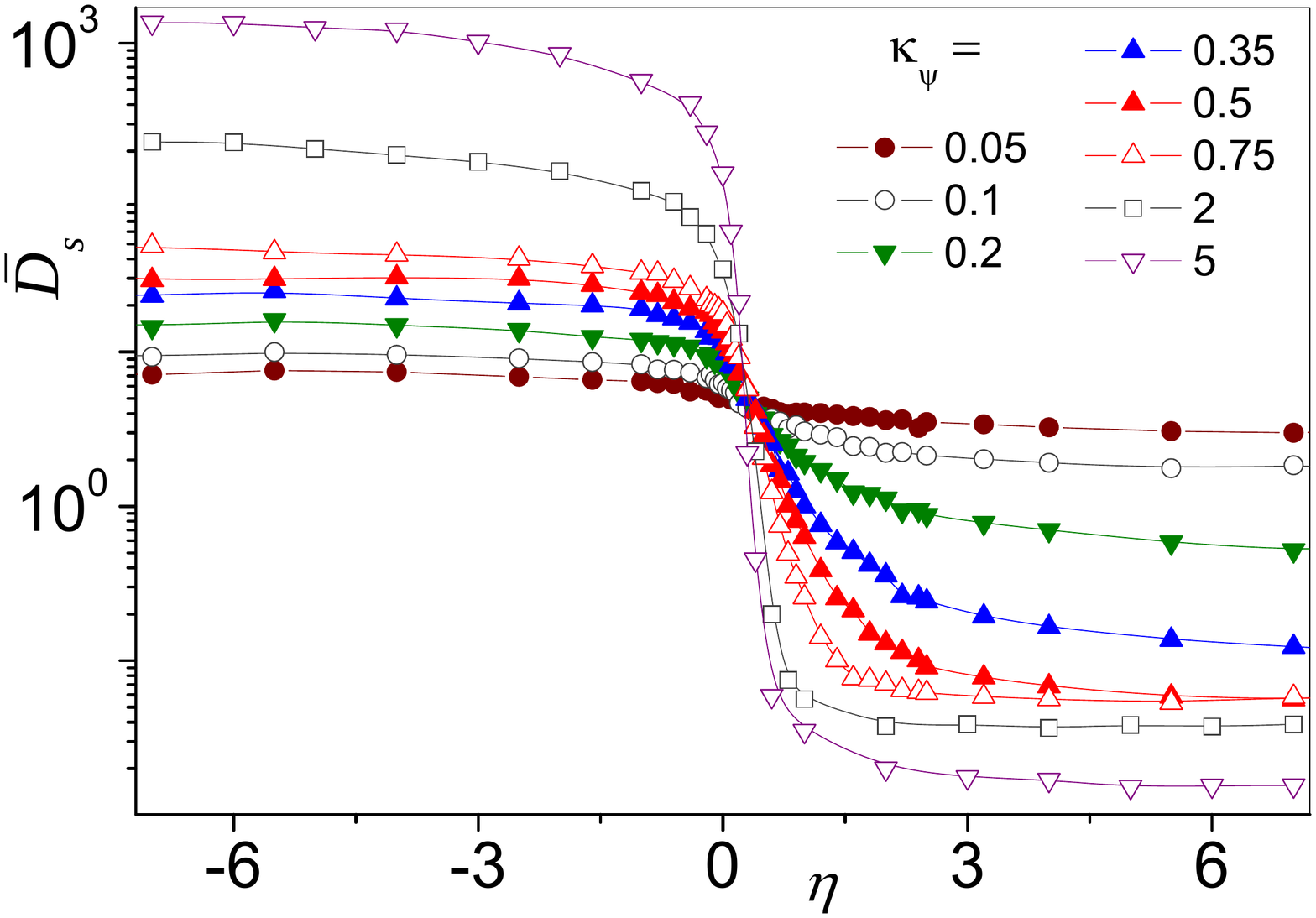}
\caption{(Color online) Diffusion dependence on towing configuration: $\bar D_s$ vs. $\eta$ for different $\kappa_\psi$. Other simulation parameters are: $D_0=0.01$, $D_\phi=0.003$, $D_\psi=0.1$, and $v_0=10$. \label{F2}}
\end{figure}

Our analysis focuses on the diffusion of the active dimer of Eq. (\ref{modelLE}) in the bulk, that is in the absence of obstacles or other geometric constraints. Numerically, the effective diffusion constant is defined through the asymptotic law \cite{Gardiner,EPJST},
$D=\lim_{t\to \infty} \langle {\vec r}^2(t)\rangle/4t$. $D$ is the sum of a constant equilibrium translational contribution, $D_0$, and a self-propulsion term, $\bar D_s$, which strongly depends on the model parameters \cite{soft,ourJCP}. Analytically, $D$ can be calculated by having recourse to Kubo's formula
\begin{eqnarray} \label{Kubo}
D=\int_0^\infty \langle \dot x(t) \dot x(0) \rangle dt=
D_0+\int_0^\infty C(t) dt,
\end{eqnarray}
where $C(t)=v_0^2\langle
\cos[\phi(t)+\psi(t)]\cos[\phi(0)+\psi(0)]\rangle$ and $D_0$ was introduced in Eq. (\ref{modelLE}). Note that our
model is isotropic, so that $D$ can equivalently be computed along
either orthogonal axis in the plane. By inspecting Eq. (\ref{Kubo}), one concludes immediately that $\bar D_s=D-D_0$, is invariant under the transformations
$\psi \to -\psi$, corresponding to inverting chirality, and $(\psi,\eta)\to (\psi+\pi, -\eta)$, or, equivalently, $U(\psi)\to -U(\psi)$. In particular, due to the latter property, it is clear that for $\eta=0$ towing by either pulling or pushing yields the same value of $\bar D_s$ \cite{ourJCP}.

In the best investigated case of an active head acting as a puller or a pusher, i.e., for $\psi(t)\equiv 0$ or $\psi(t)\equiv \pi$ ({\it dumbbell} approximation \cite{GrahamPRL}), the angular noise strength, $D_\phi$, plays the role of an orientational diffusion constant, whose inverse, $\tau_\phi$, quantifies the temporal persistency of the ensuing  active Brownian motion \cite{EPJST}. For long observation times $t$, $t\gg \tau_\phi$, the definition of the effective
diffusion constant in Eq, (\ref{Kubo}) yields $\bar D_s=D_s$, where $D_s=v_0^2/2D_\phi$ is the
self-propulsion term obtained by eliminating the angular variable $\psi$ from the model Eqs. (\ref{modelLE}) \cite{memory,ourJCP}.

\begin{figure}[tp]
\centering
\includegraphics[width=7.5cm]{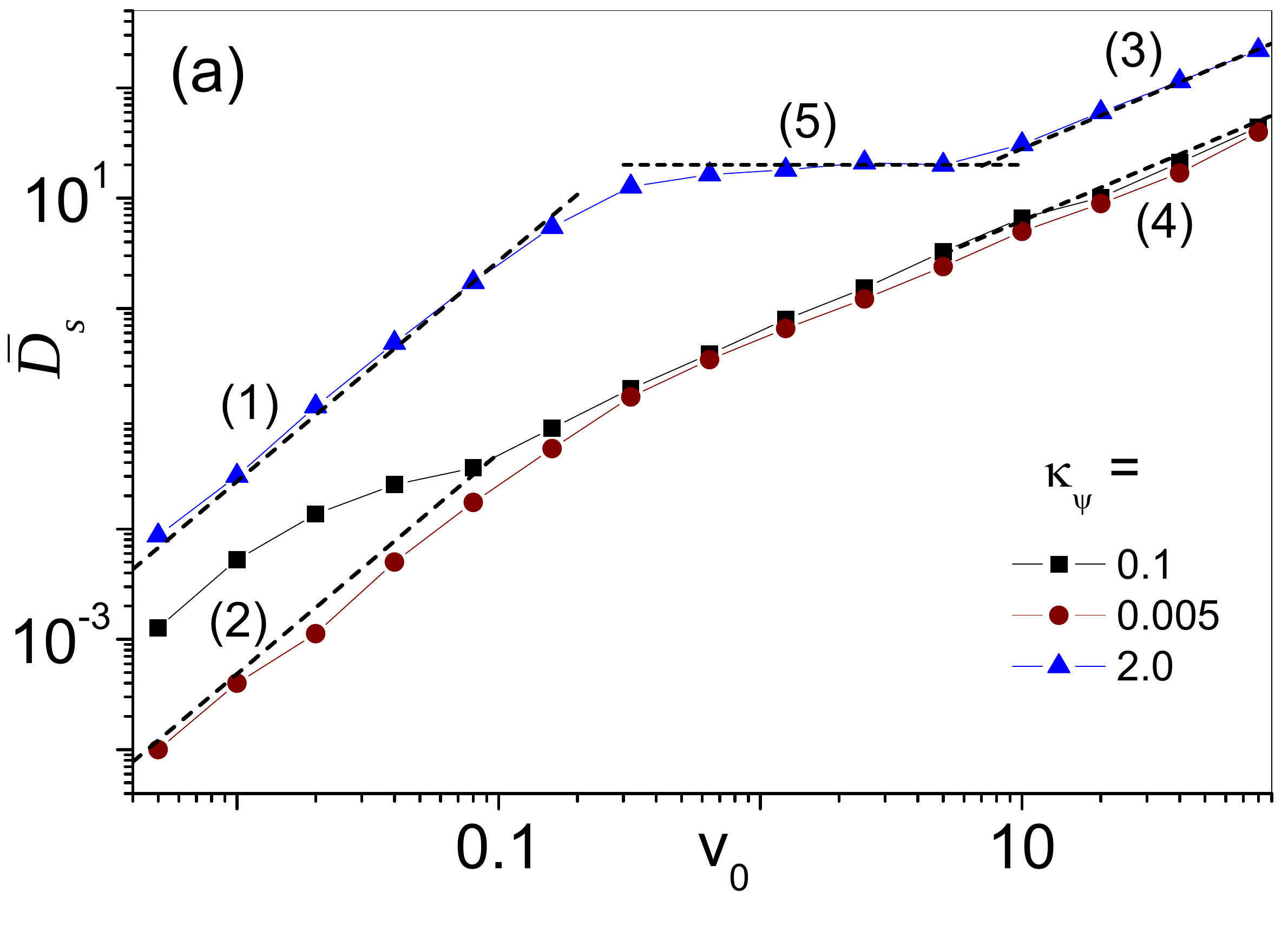}
\includegraphics[width=7.5cm]{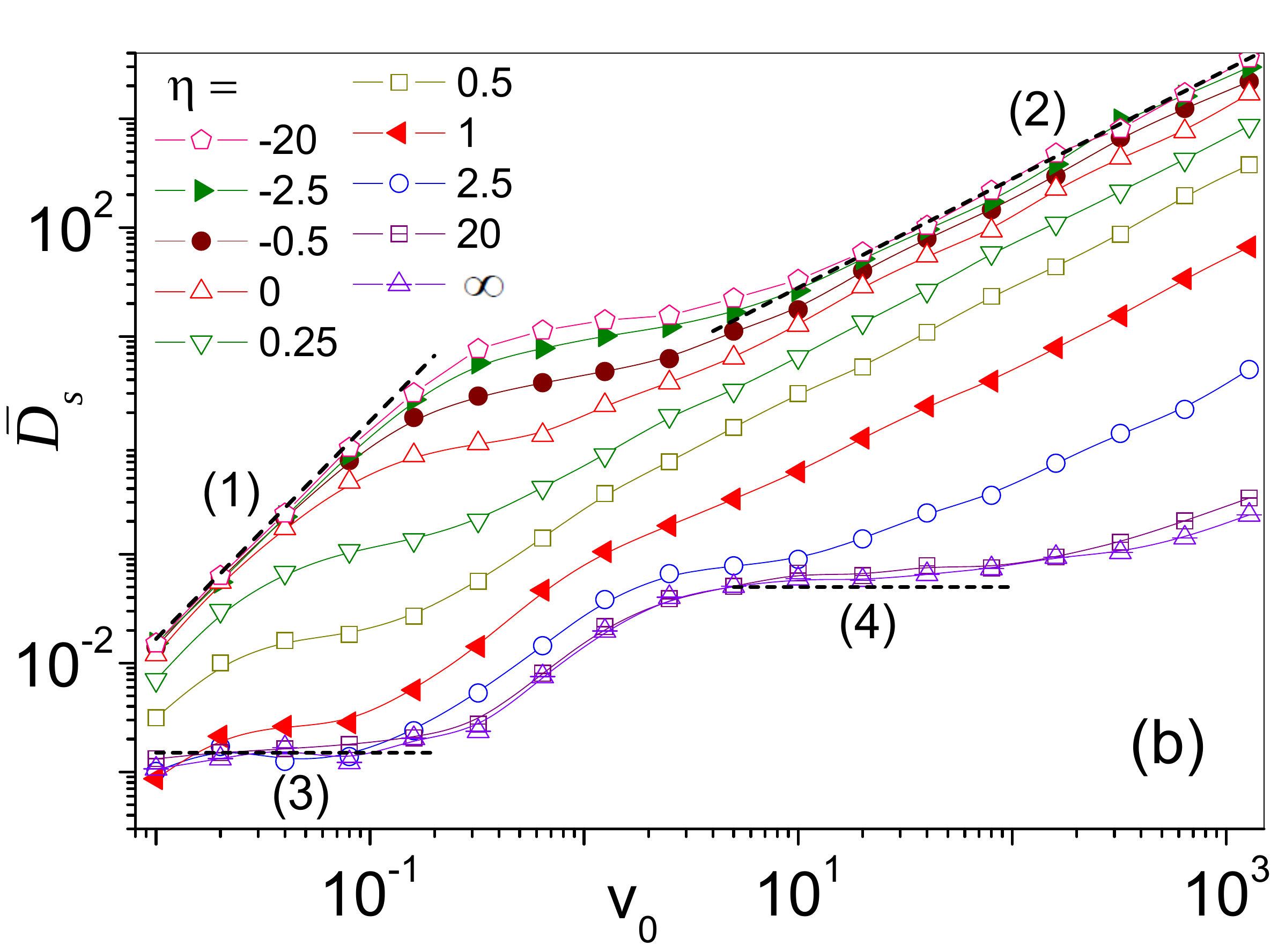}
\caption{(Color online) Diffusion dependence on dimer's self-propulsion speed: $\bar D_s$ vs. $v_0$ for (a) $\eta=0$ and different $\kappa_\psi$, (b) $\kappa_\psi=0.5$ and different $\eta$. Other simulation parameters are: $D_0=0.01$, $D_\phi=0.003$, and $D_\psi=0.1$. The dashed lines in (a) are analytical estimates obtained from Ref. [16]: (1) quadratic regime of Eq. (\ref{quad2}); (2) quadratic regime of Eq. (\ref{quad1}); (3) linear regime of Eq. (\ref{lin1}); (4) linear regime of Eq. (\ref{lin2}); and (5) plateau of Eq. (\ref{plateau}). The dashed lines in (b) represent the analytical estimates discussed in the text: (1) $\bar D_s=D_s$; (2) $\bar D_s$ as in Eq. (\ref{lin1}) but for $\kappa_\psi \to 4\kappa_\psi$; (3) $\bar D_s=D_\phi/2$; and (4) $\bar D_s=D_\psi/2$.
\label{F3}}
\end{figure}

The stochastic differential Eqs. (\ref{modelLE}) were numerically
integrated by means of a standard Euler-Maruyama scheme
\cite{Kloeden}. The stochastic averages were taken over ensembles of
up to 10$^5$ trajectories with random initial conditions of the angular variables $\phi$ and $\psi$. All trajectories have been sampled for time intervals long enough to ignore transient effects.

\begin{figure}[tp]
\centering
\includegraphics[width=7.0cm]{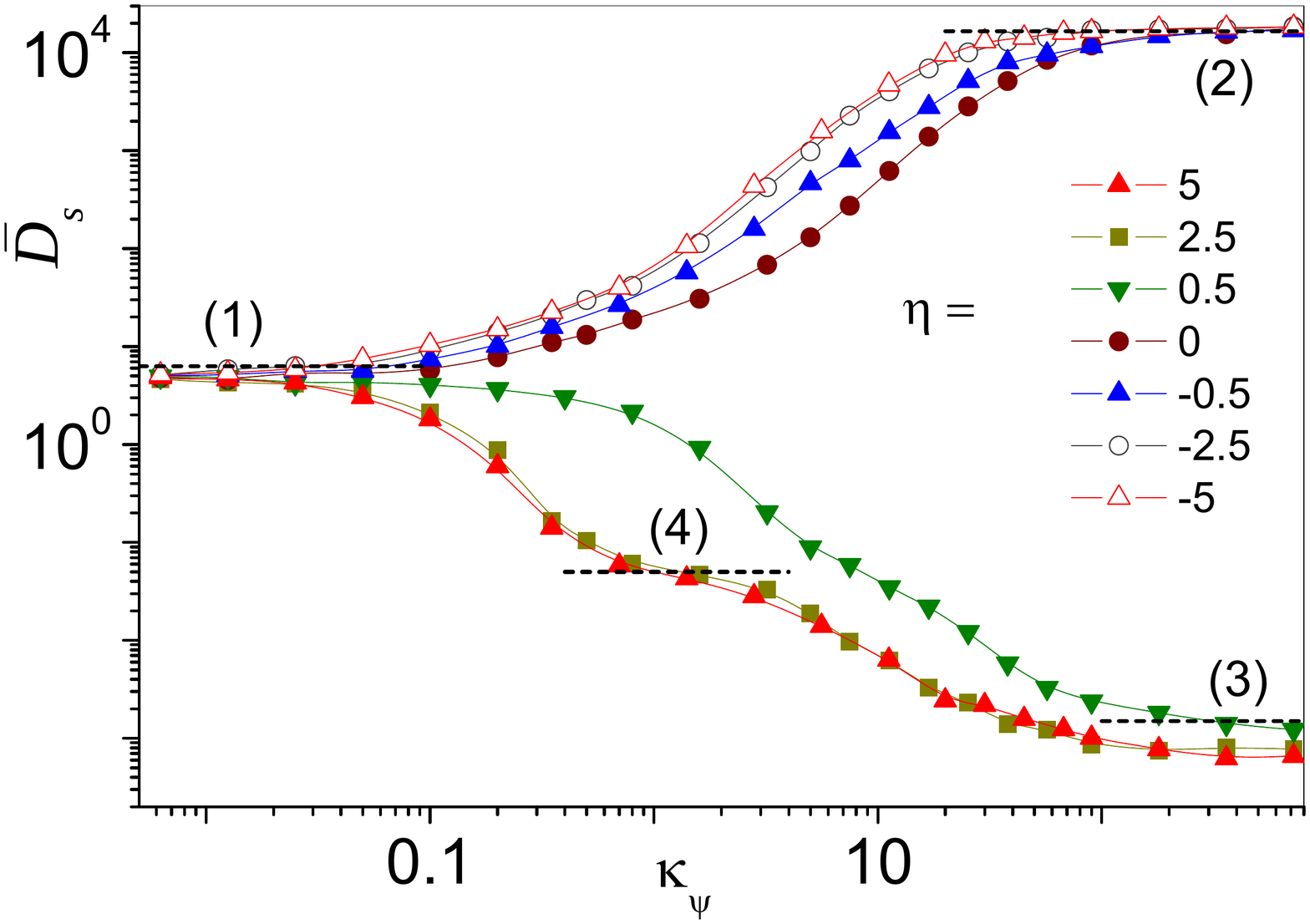}
\caption{(Color online) Diffusion dependence on the tower-cargo coupling: $\bar D_s$ vs. $\kappa_\psi$ for different $\eta$. Other simulation parameters are: $D_0=0.01$, $D_\phi=0.003$, $D_\psi=0.1$, and $v_0=10$. The dashed lines are the asymptotes introduced in the text: (1) $\bar D_s$ of Eq. (\ref{lin2}); (2) $\bar D_s=D_s$; (3) $\bar D_s=D_\phi/2$; and (4) $\bar D_s=D_\psi/2$.
\label{F4}}
\end{figure}

Numerical simulation shows that, in general, the diffusion constant of an active dimer strongly depends on the details of its orientational dynamics. Our numerical data are displayed in Figs. \ref{F2}-\ref{F4}, where the self-propulsion term of the diffusion constant, $\bar D_s$, is plotted for different choices of the model parameters. A few qualitative properties are apparent by inspection:

(1) {\it $\eta$ dependence}. In Fig. \ref{F2} the dependence of $\bar D_s$ on the towing configuration is manifest. For $\eta=0$ and $\eta\to - \infty$ the angular potential, $U(\psi)$, approaches a sinusoidal profile, respectively with one stable pulling configuration, $\psi_0=0$, and two bistable configurations, one pulling, $\psi_0=0$, and one pushing, $\psi_0=\pi$. In both cases we obtain relatively large values of $\bar D_s$. However, for $\eta \to -\infty$, both stable configurations equally contribute to the dimer diffusion and, therefore, $\bar D_s$ is larger than for $\eta=0$. Vice versa, in the limit $\eta \to \infty$ the dimer enters a stalled configuration with $\psi_0=\pm \pi/2$. The ensuing chiral dynamics is known to be characterized by reduced diffusivity \cite{EPJST,Lowen}, whence the dramatic drop of $\bar D_s$ for $\eta \gg 1/4$, reported in figure. As expected, such an $\eta$-dependence is enhanced by increasing the coupling constant $\kappa_\psi$, see item (3) below.

(2) {\it $v_0$ dependence}. In Fig. \ref{F3} we illustrate the dependence of $\bar D_s$ on the JP drive for different values of $\kappa_\psi$, panel (a), and $\eta$, panel (b). We know from Ref. [16] that the dumbbell approximation applies only for low $v_0$ {\it and} large $U(\psi)$ amplitudes. Indeed, the condition of sufficiently slow self-propulsion speed alone does ensure a quadratic dependence of $\bar D_s$ on $v_0$; however,  the ratio $\bar D_s/D_s$ grows asymptotically to $1$ solely in the limit $\kappa_\psi \to \infty$. These predictions are confirmed by the curves plotted in Fig. \ref{F3}(a). Finally, on increasing the self-propulsion speed, $\bar D_s$ eventually develops a peculiar linear dependence on $v_0$.
Such a dependence is typical of structured non-rigid swimmers, where the swimmer's orientation follows with finite time delay the fluctuating self-propulsion velocity of its active component \cite{soft}. Moreover, the quadratic and linear branches of the curves $\bar D_s$ versus $v_0$ are separated by a single plateau for negative $\eta$, and by two distinct plateaus for (strongly) positive $\eta$, see Figs. \ref{F3}(a) and (b). The huge diffusion drop observed in Fig. \ref{F2}, as $\eta$ increases from negative to positive values, is corroborated by the data sets of Fig. \ref{F3}(b). Note that this statement holds as long as $v_0$ is not vanishingly small. [For $v_0\to 0$ our data suggest that $\bar D_s \to D_s$, no matter what $\eta$.]

(3) {\it $\kappa_\psi$ dependence}. As anticipated in item (1), the drop of $\bar D_s$ with increasing $\eta$ is magnified at high coupling constants. Indeed, raising the $U(\psi)$ amplitude makes the different propulsion configurations more stable against angular fluctuations, thus enhancing their distinct impact on the dimer's diffusive dynamics. In Fig. \ref{F4} such a behavior is manifest for $\kappa_\psi \to \infty$; vice versa, all displayed data sets seem to converge toward a unique horizontal asymptote for $\kappa_\psi \to 0$.

We now interpret the above diffusion properties by applying the analytical approach of Refs. [16,27], based on Kubo's formula (\ref{Kubo}). Let us start with the case $\eta=0$.  The dynamics of an active dimer with a pulling head is easily amenable to that of an eccentric swimmer. In Sec. 4 of Ref. [16] the quadratic dependence of $\bar D_s$ on $v_0$ has been explicitly calculated  for $v_0\ll \kappa_\psi$, namely,
\begin{equation}
\label{quad1}
\bar D_s=D_s/(D_\psi/D_\phi+1)= v_0^2/[2(D_\phi+D_\psi)],
\end{equation}
for $\kappa_\psi\ll D_\phi$, and
\begin{equation}
\label{quad2}
\bar D_s=D_s\;e^{-D_\psi/\kappa_\psi},
\end{equation}
for $\kappa_\psi \gg D_\phi$. The linear dependence of $\bar D_s$ on $v_0$, instead, sets in for $v_0\gg \kappa_\psi$, and can be approximated to  \cite{soft},
\begin{equation}
\label{lin1}
\bar D_s=(v_0/2)\sqrt{\pi/2}\sqrt{\kappa_\psi/D_\psi},
\end{equation}
for $\kappa_\psi\gg D_\psi$, and
\begin{equation}
\label{lin2}
\bar D_s=(v_0/2)\sqrt{\pi/2},
\end{equation}
for $\kappa_\psi \ll D_\psi$. All four analytical regimes are clearly discernible in Fig. \ref{F3}(a).

In Fig. \ref{F4}, the unique limit of the curves $\bar D_s$ for $\kappa_\psi /v_0 \to 0$ is well reproduced by the linear regime of Eq. (\ref{lin2}). In the opposite limit, $\kappa_\psi/v_0 \to \infty$, two distinct horizontal asymptotes emerge for $\eta \to \pm \infty$. On recalling the dumbbell approximation, no surprise that under the conditions, $\eta \to -\infty$ and $\kappa_\psi/v_0 \to \infty$, one recovers $\bar D_s=D_s$. The conditions  $\eta \to \infty$ and $\kappa_\psi/v_0 \to \infty$, instead, define the diffusion dynamics of a chiral dumbbell subject to the effective torques $\Omega=\pm v_0$ (in correspondence with the $U$ minima $\psi_0=\mp \pi/2$). The diffusion of a chiral swimmer obeys the simple law \cite{EPJST,Lowen},  $\bar D_s=D_s/[1+(\Omega/D_\phi)^2]$. In Fig. \ref{F4}, $v_0 \gg D_\phi$, and therefore $\bar D_s \simeq D_\phi/2$.

A plateau separates the quadratic, Eq. (\ref{quad1}), from the linear, Eq. (\ref{lin1}), branch of $\bar D_s$ for $\eta=0$ and $\kappa_\psi \gg D_\psi$. For $(v_0/\kappa_\psi)^2 \gg D_\phi/D_\psi$ an analytical estimate yields \cite{soft},
\begin{equation}
\label{plateau}
\bar D_s=\kappa_\psi^2/2D_\psi,
\end{equation}
in agreement with the numerical data plotted in Figs. \ref{F3} and \ref{F4}. Upon varying $\eta$ the plateau structure changes. By decreasing $\eta$ from $0$ down to $-\infty$, the profile of $U(\psi)$ changes from $-\kappa_\psi\cos \psi$  to $-\kappa_\psi\cos 2\psi$. Provided that the fluctuations around the minima $\psi_0=0$ and $\pi$ are small, i.e., $\kappa_\psi \gg D_\psi$, this amounts to replacing $\kappa_\psi$, for $\eta=0$, with $4\kappa_\psi$, for $\eta \to - \infty$. Accordingly, the curve $\bar D_s$ versus $v_0$ for $\eta \to -\infty$ overlaps the corresponding curve at $\eta=0$ in the quadratic regime of Eq. (\ref{quad1}), is about 16 times larger at the plateau, Eq. (\ref{plateau}), and only 4 times steeper in the linear regime of Eq. (\ref{lin1}). For $\eta \to \infty$, the minimum, $\psi_0=0$, of the angular potential splits into a pair of bistable minima, $\psi_0=\pm \pi/2$. The diffusion of the {\it chiral} dimer is then suppressed and, for $v_0\ll \kappa_\psi$, one recovers the plateau $\bar D_s=D_\phi/2$, already
detected in Fig. \ref{F4}. The second plateau that appears on further increasing $v_0$ is detectable only at large $\eta$ values under the conditions $D_\psi \ll 4 \kappa_\psi$ and $v_0 <4\kappa_\psi$, see Fig. \ref{F4}. This situation is reminiscent of the linear diffusion regime of chiral eccentric swimmers analyzed in Sec. 5 of Ref. [16]. Accordingly, the expected value of the plateau, in the present units, reads
$\bar D_s\simeq D_\psi/2$. The plateau structure of $\bar D_s$ versus $v_0$ is well illustrated in Fig. \ref{F3}(b).

In conclusion, the diffusivity of a tightly bound active dimer is strongly influenced by short-range hydrodynamic interactions that tend to orient its active and passive components one relative to the other \cite{Baraban}. An active microswimmer coupled to a cargo exhibits optimal towing capability only in the puller and pusher configurations, whereas in all other configurations it drives a chiral dynamics, which suppresses the motility of the swimmer-cargo system. The present analysis can be easily extended to the more general case of cargoes of large dimensions, namely to asymmetric active dimers, by accounting for long-range hydrodynamic interactions \cite{Ermak} purportedly ignored in the present report.

\section*{Acknowledgements}
We thank RIKEN's RICC for computational resources.
Y. Li is supported by the NSF China under grant No. 11505128 and the Tongji University under grant No. 2013KJ025.
P.K.G. is supported  by SERB Start-up Research Grant (Young Scientist) No. YSS/2014/000853 and the UGC-BSR Start-Up Grant No. F.30-92/2015

\end{document}